\documentclass[aps,pre,showpacs]{revtex4-1}

\usepackage{amsfonts}
\usepackage{bbm}
\usepackage{array}
\usepackage{amsmath}
\usepackage{mathrsfs}
\usepackage{mathtools}
\usepackage{amssymb}
\usepackage{upgreek}
\usepackage{yhmath}
\usepackage{graphicx}
\begin{document}


\title{A {S}chur decomposition reveals the richness of structure in homogeneous, isotropic turbulence as a consequence of localised shear.}


\author{C. J. Keylock}
\affiliation{Department of Civil and Structural Engineering and Sheffield Fluid Mechanics Group, University of Sheffield, Sheffield, S1 3JD, U.K.\\}


\date{\today}

\begin{abstract}
An improved understanding of turbulence is essential for the effective modelling and control of industrial and geophysical processes. Homogeneous, isotropic turbulence (HIT) is the archetypal field for developing turbulence physics theory. Based on the Schur transform, we introduce an additive decomposition of the velocity gradient tensor into a normal part (containing the eigenvalues) and a non-normal or shear-related tensor. We re-interrogate some key properties of HIT and show that the the tendency of the flow to form disc-like structures is not a property of the normal tensor; it emerges from an interaction with the non-normality. Also, the alignment between the vorticity vector and the second eigenvector of the strain tensor is another consequence of local shear processes.
\end{abstract}

\pacs{47.27.Ak, 47.27.Gs, 47.32.C-}


\maketitle


\section{Introduction}
A great range of physical phenomena, from magnetohydrodynamic systems, quantised vortices in liquid Helium and solar plasmas, to drag reduction on an airfoil and the flight of a bumble bee require, for their understanding, an engagement with the physics of turbulence - the mechanisms by which energy produced at large scales is transferred to, and dissipated, at small scales \cite{arneodo96,k16b,roux16}. For classical turbulence, continuity and the Navier-Stokes equations govern the nature of the problem, but do not permit us to infer the mechanisms for interscale transfers, leading to different hypotheses for the relevant scaling laws for dissipation \cite{K41,vassilicos15} and attempts to relate flow topology to the mechanisms of the energy cascade \cite{goto08}. 

A key property of turbulence is the velocity gradient tensor,  
\begin{equation}
\mathsf{A} = \frac{\partial u_{i}}{\partial x_{j}} \equiv \biggl(\begin{smallmatrix}
\partial u_{1}/\partial x_{1} & \partial u_{1}/\partial x_{2} & \partial u_{1}/ \partial x_{3}\\ 
\partial u_{2}/\partial x_{1} & \partial u_{2}/\partial x_{2} & \partial u_{2}/ \partial x_{3}\\ 
\partial u_{3}/\partial x_{1} & \partial u_{3}/\partial x_{2} & \partial u_{3}/ \partial x_{3}
\end{smallmatrix} \biggr),
\end{equation}
where $x$ is a length, $u$ is a velocity component, and the subscripts indicate the relevant, orthogonal orientation. It can be shown that $\mathsf{A}$ is intimately associated with an understanding of the incompressible Navier-Stokes equations by taking their spatial gradient:
\begin{equation}
\frac{\partial}{\partial t}\mathsf{A} + \mathsf{u} \cdot \nabla \mathsf{A} = -\mathsf{A}^{2} - \mathsf{H} + \nu \nabla^{2} \mathsf{A},
\label{eq.NS}
\end{equation}
where $\nu$ is the kinematic viscosity, $\mathsf{H}$ is the Hessian of the kinematic pressure field, i.e. $H_{ij} = \frac{\partial^{2} p}{\partial x_{i} \partial x_{j}}$, with $p$ denoting the pressure. 

Since the development of accurate numerical analyses of homogeneous, isotropic turbulence in the early-1980s \cite{kerr85}, there have been a number of discoveries regarding properties of the velocity gradient tensor, $\mathsf{A}$, of homogeneous, isotropic turbulence, many of which were summarized in the review paper by Meneveau \cite{meneveau11}. In this paper, we re-examine two of these key results: the structure of the eigenvalues of the strain rate tensor, and the alignment between the strain rate tensor and the vorticity vector. In addition to these properties, although it is not the topic of direct investigation in this paper, several models have been developed for the dynamics of turbulence by formulating ordinary differential equations describing the evolution of properties of $\mathsf{A}$ \cite{cantwell92,chevillard08,wilczek14}. However, little emphasis has been placed on the role of small-scale shear in the flow. In part, this is because typical approaches to analysis, such as the study of the second and third invariants and the characteristic equation for $\mathsf{A}$ (the $Q-R$ diagram \cite{tsinober01,laizet15}), place emphasis on scalar quantities that are derived from the eigenvalues of $\mathsf{A}$, which are not related to shear. In part, it is because the standard decomposition of the velocity gradient tensor is into strain, $\mathsf{S}$, and rotation, $\mathsf{\Omega}$, components, which underplays the role of shear, whose effect is divided across both these tensors: 
\begin{eqnarray}
\mathsf{A} &=& \mathsf{S} + \mathsf{\Omega} \nonumber \\
\mathsf{\Omega} &=& \frac{1}{2}(\mathsf{A} - \mathsf{A}^{*}) \nonumber \\
\mathsf{S} &=& \frac{1}{2}(\mathsf{A} + \mathsf{A}^{*}),
\end{eqnarray}
and the asterisk denotes the conjugate transpose. Consider, by way of example, an hypothetical shear acting at the scale of the tensor given by $\mathsf{A} = \biggl(\begin{smallmatrix}
0 & 0 & 0\\ 0 & 0 & 0\\2 & 0 & 0
\end{smallmatrix} \biggr)$. Clearly, the $i = 1, \ldots, 3$ eigenvalues of $\mathsf{A}$, $e_{i}$ are all zero and therefore, the trace-free constraint for incompressibility is maintained. However, this means that the invariants of the characteristic equation for the tensor, which may be determined from these eigenvalues, are also all zero. With the characteristic equation given by
\begin{equation}
e_{i}^{3} + P e_{i}^{2} + Q e_{i} + R = 0,
\end{equation}
the invariants are
\begin{eqnarray}
P &=& \sum e_{i} \nonumber \\
Q &=& \sum (1 - \delta_{ij}) e_{i}e_{j} \nonumber \\
R &=& - \prod e_{i}, 
\label{eq.PQR}
\end{eqnarray}
where $P = 0$ for incompressibility. This means that the discriminant function separating regions with real eigenvalues to those with a conjugate pair and a real value, simplifies to
\begin{equation}
\Delta = Q^{3} + \frac{27}{4}R^{2}.
\label{eq.Delta}
\end{equation}

With all of these terms zero for any pure shear such as that in the example case, it is clear that the role of shear is not revealed in this approach to studying $\mathsf{A}$. Furthermore, for our example tensor, $\mathsf{S} = \biggl(\begin{smallmatrix}
0 & 0 & 1\\ 0 & 0 & 0\\1 & 0 & 0
\end{smallmatrix} \biggr)$ and $\mathsf{\Omega} = \biggl(\begin{smallmatrix}
0 & 0 & -1\\ 0 & 0 & 0\\1 & 0 & 0
\end{smallmatrix} \biggr)$, meaning that shear effects are not isolated within a joint Hermitian-skew Hermitian (strain-rotation) decomposition. Consequently, the commonly considered restricted Euler approximation for the dynamics of $\mathsf{A}$, which focuses on the second ($Q$) and third ($R$) invariants \cite{vieillefosse84,cantwell92}, or work that tries to understand turbulence structure in relation to $Q$ and $R$ \cite{tsinober01} eliminates shear from explicit consideration. The focus of this paper is to develop and employ a different, additive decomposition of the velocity gradient tensor. There are two reasons to do this: First, we can isolate eigenvalue-related phenomena in one tensor and non-normal, shearing effects in the other, making explicit the role of small-scale shearing between vortices, for example. Second, we can extract the strain and rotation tensors, for the normal and non-normal tensors, providing a new means for studying properties of the velocity gradient tensor. 

\section{An additive decomposition that highlights shear effects}
The Schur decomposition \cite{schur1909} provides a means for obtaining an additive shear-based decomposition of $\mathsf{A}$:
\begin{equation}
\mathsf{A} = [\mathsf{U}\mathsf{T}\mathsf{U}^{*}]_{\mathbb{C}} = [\mathsf{U}(\mathsf{L} + \mathsf{N})\mathsf{U}^{*}]_{\mathbb{C}},
\label{eq.schur}
\end{equation} 
where the subscript after the square brackets indicates either a real, $\mathbb{R}$, or complex, $\mathbb{C}$, transform is being undertaken, $\mathsf{U}$ is unitary, and the Schur tensor, $[\mathsf{T}]_{\mathbb{C}}$ is formed additively from a diagonal matrix of eigenvalues, $\mathsf{L}$ and a matrix of non-normality, $\mathsf{N}$. This is upper triangular for the complex transform, and upper triangular for the real form if the eigenvalues are real. Some complexity in our exposition, below, results from the fact that the real transform is only quasi-upper triangular if the eigenvalues are complex (the conjugate pair form a $2 \times 2$ Jordan block, with the off-diagonal terms integrating information on the imaginary part of the eigenvalues and some of the non-normality). The Schur transform is not unique because the ordering of the eigenvalues along the diagonal of $\mathsf{T}$ is not set. However, given an arbitrary ordering it is possible to rotate the transform into a standardised arrangement, and in this work we consider that $L_{11}$ contains the most negative eigenvalue, $e_{1}$, of $\mathsf{A}$, with $L_{33}$ the most positive, $e_{3}$.

If adopting the complex Schur decomposition, the approach is simple to state:
\begin{equation}
\mathsf{A} = \mathsf{B} + \mathsf{C},
\label{eq.additive}
\end{equation}
where $\mathsf{B} = [\mathsf{U}\mathsf{L}\mathsf{U}^{*}]_{\mathbb{C}}$ and $\mathsf{C} = [\mathsf{U}\mathsf{N}\mathsf{U}^{*}]_{\mathbb{C}}$. This approach is additive and energy-preserving, although it may yield complex-valued tensors $\mathsf{B}$ and $\mathsf{C}$, where $\mbox{Im}(B_{ij}) = -\mbox{Im}(C_{ij})$ to maintain additivity. If $||\mathsf{N}||_{F} = 0$, where $||\ldots||_{F}$ is the Frobenius norm, then $\mathsf{T} = \mathsf{L}$ and $\mathsf{U}$ may be interpreted as the set of eigenvectors corresponding to $\mathsf{L}$. In this situation, $\mathsf{B} = \mathsf{U}\mathsf{L}\mathsf{U}^{*}$ is equal to $\mathsf{A}$ and $\mathsf{C} = \varnothing$.

An alternative approach is to enforce that $\mathsf{B}$ and $\mathsf{C}$ are real. This requires a slight modification to the above approach when $e_{i} \in \mathbb{C}$. Using the real Schur transform, $\mathsf{A} = [\mathsf{U}\mathsf{T}\mathsf{U}^{*}]_{\mathbb{R}}$, there are two relevant cases:
\begin{enumerate}
\item There is an excess of total strain over enstrophy ($\Delta < 0$), meaning that all the eigenvalues are real. In which case, $[\mathsf{N}]_{\mathbb{R}}$ is upper triangular and we simply have $\mathsf{B} = [\mathsf{U}\mathsf{L}\mathsf{U}]_{\mathbb{R}}^{*}$ and $\mathsf{C} = [\mathsf{U}\mathsf{N}\mathsf{U}]_{\mathbb{R}}^{*}$;
\item For $\Delta > 0$, there will be a conjugate pair of eigenvalues for $\mathsf{A}$ and, as noted above, $[\mathsf{T}]_{\mathbb{R}}$ is then only quasi-upper triangular. In this case, we have $\mathsf{B} = [\mathsf{U}]_{\mathbb{R}} \tilde{\mathsf{L}} [\mathsf{U}]_{\mathbb{R}}^{*}$, where 
\begin{eqnarray}
\tilde{\mathsf{L}} &=& \biggl(\begin{smallmatrix}
[L_{11}]_{\mathbb{R}} & \pm \mbox{Im}([L_{11}]_{\mathbb{C}}) & 0\\ 
\mp \mbox{Im}([L_{11}]_{\mathbb{C}}) & [L_{22}]_{\mathbb{R}} & 0\\ 
0 & 0 & [L_{33}]_{\mathbb{R}} \end{smallmatrix} \biggr), \,\,\,\mbox{or}\,\,\, \nonumber \\
&=& \biggl(\begin{smallmatrix}
[L_{11}]_{\mathbb{R}} & 0 & 0\\ 
0 & [L_{22}]_{\mathbb{R}} & \pm \mbox{Im}([L_{33}]_{\mathbb{C}})\\
 0 & \mp \mbox{Im}([L_{33}]_{\mathbb{C}}) & [L_{33}]_{\mathbb{R}} \end{smallmatrix} \biggr),
\end{eqnarray}
depending on the plane where the conjugate pair is located.
\end{enumerate}
For this real decomposition (\ref{eq.additive}) holds as does the additive decomposition of total strain $||\mathsf{S}_{A}||_{F}^{2} = ||\mathsf{S}_{B}||_{F}^{2} + ||\mathsf{S}_{C}||_{F}^{2}$. However, there is an issue with preservation of the enstrophy as a consequence of missing imaginary contributions to the norm of $||\mathsf{\Omega}_{C}||_{F}$. Given the dynamical significance of enstrophy an energy conservation, we prefer to work with the complex Schur decomposition in this paper. 

As a consequence, because the eigenvalues of $\mathsf{A}$ and $\mathsf{B}$ are identical, then the second invariant, $Q_{{A}} =  Q_{{B}}$ from (\ref{eq.PQR}) and $Q_{{C}} = 0$. Hence, $||\mathsf{\Omega}_{C}||_{F} = ||\mathsf{S}_{C}||_{F}$, and rewriting $Q$ in its physically more insightful form as the excess of enstrophy over total strain
\begin{equation}
Q = \frac{1}{2}(||\mathsf{\Omega}||_{F}^{2} - ||\mathsf{S}||_{F}^{2}),
\label{eq.Q}
\end{equation}
it is clear that the enstrophy and total strain in $\mathsf{B}$ must be less than that in $\mathsf{A}$ by a constant equal to both $\mathsf{\Omega}_{C}||_{F}^{2}$ and $\mathsf{S}_{C}||_{F}^{2}$. 

\begin{table}
\caption{Properties of the HIT simulation in the Johns Hopkins database \cite{yili}.}
\label{tab.HIT}
\begin{center}
\begin{tabular}{lc}
\hline
 Property & Value\\
\hline 
Grid & $1024^{3}$ periodic box\\
Domain & $[0, 2\pi]^{3}$\\
Viscosity, $\nu$ & $1.85 \times 10^{-4}$\\
Mean dissipation rate, $\epsilon$ & 0.0928\\
Taylor micro-scale, $\lambda$ & 0.118\\
Taylor Reynolds number, $Re_{\lambda}$ & 433\\
Kolmogorov length, $\eta$ & $2.87 \times 10^{-3}$\\
\hline
\end{tabular}
\end{center} 
\end{table}

\section{Data for homogeneous, isotropic turbulence}
Our intention in this paper is to highlight the role of local shear (at the scale of the velocity gradient tensor) for influencing the statistical properties of turbulence in the absence of mean shear. Our Schur-based decompositions isolate this effect in $\mathsf{N}$. To demonstrate the significance of non-normality/local shear, while avoiding mean shear, we focus on homogeneous, isotropic turbulence (HIT), and we make use of the Johns Hopkins numerical simulation of HIT at a Taylor Reynolds number of 433 \cite{yili}, which has emerged as an important resource for studying turbulence \cite{luthi09,wan2010,lawsondawson15}. Further details on the properties of this simulation are given in Table \ref{tab.HIT}. We resolved the flow at a spatial sampling interval of $0.052 \lambda$ and studied the spatial properties at a particular sampling instance.  

\section{Results}

\begin{figure}
\vspace*{2mm}
\begin{center}
\includegraphics[width=8.8cm]{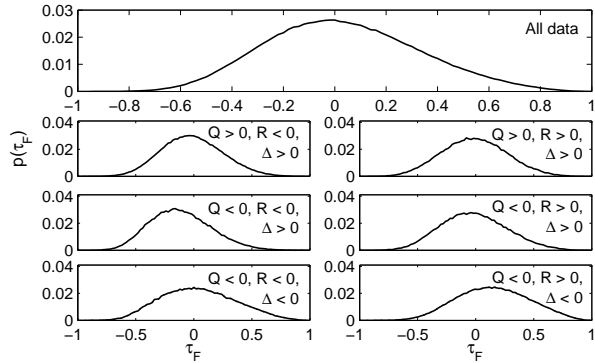}
\end{center}
\caption{The distribution functions for the Frobenius norm ratio, $\tau_{F}$, shown for all the data and as a function of the six regions of the $Q-R$ diagram in the various panels.} 
\label{fig.frobnorm}
\end{figure}

\subsection{Frobenius norms of the tensors, $\mathsf{B}$ and $\mathsf{C}$}
If non-normality was not significant to the behaviour of homogeneous, isotropic turbulence, then the ratio of the Frobenius norms of $\mathsf{B}$ and $\mathsf{C}$ should strongly favour $\mathsf{B}$. We define the ratio
\begin{equation}
\tau_{f} = \frac{(||\mathsf{B}||_{F} - ||\mathsf{C}||_{F})}{(||\mathsf{B}||_{F} + ||\mathsf{C}||_{F})}.
\label{eq.tauF}
\end{equation}
The upper panel of Fig. \ref{fig.frobnorm} shows that the mode of the distribution function is slightly negative and the median is very close to $\tau_{F} = 0$. Hence, the non-normality is as important as the normal part explained by the eigenvalues. Consequently, our first result is that local shear is of significance to the dynamics of even homogeneous isotropic turbulence. When the results in Fig. \ref{fig.frobnorm} are partitioned by the six regions of the $Q-R$ diagram, we see that when $R < 0$ (left-hand column) the distribution shows greater $\mathsf{C}$ dominance than when $R > 0$. One can rewrite the expression for $R$ in (\ref{eq.PQR}) in terms of the excess of strain production and enstrophy production in a similar spirit to (\ref{eq.Q}):
\begin{equation}
R = -\mbox{Det}(\mathsf{S}) - \boldsymbol{\omega}^{*}\mathsf{S}\boldsymbol{\omega},
\label{eq.R}
\end{equation}
where $\boldsymbol{\omega}$ is the vorticity vector and both terms are typically positive \cite{betchov56}. Thus, our second result is that when $R<0$ and enstrophy production dominates, there is a greater probability for shear effects to be greater in magnitude than the eigenvalue-associated effects that are related to $\mathsf{B}$. Relative to the global data, the regions of the $Q-R$ diagram that exhibit the greatest differences are the $\Delta<0, R>0$ region, where straining and strain production are dominating, and where shear is relatively subdued, and in the narrow region where $R<0$, $Q<0$ but $\Delta>0$ and we have complex eigenvalues to $\mathsf{A}$, but total strain still exceeds enstrophy. In this region, non-normality clearly dominates with 68\% of the distribution function's mass in the $-\tau_{F}$ region. Conversely, only $1/3$ of the mass is in the $-\tau_{F}$ region when $\Delta<0$ and $R>0$. 

\begin{figure}
\vspace*{2mm}
\begin{center}
\includegraphics[width=8.8cm]{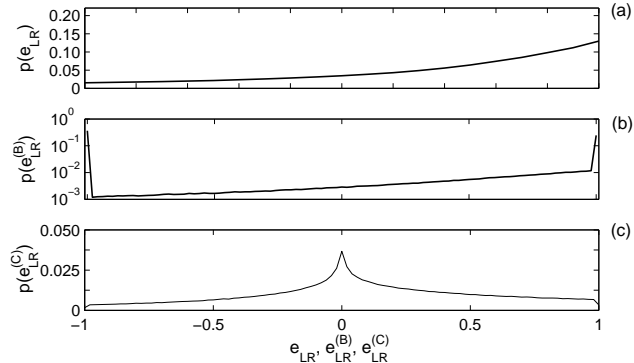}
\end{center}
\caption{The distribution function for $e_{LR}$ for HIT is shown in panel (a). Panels (b) and (c) show the corresponding distribution functions for $e_{LR}^{(\mathsf{B})}$ and $e_{LR}^{(\mathsf{C})}$. Note the use of a log-scale on the ordinate of panel (b) as a consequence of the very strong tendency for $e_{LR}^{(B)}$ to converge on $\pm 1$. These results do not depend on the use of complex or real Schur transforms.} 
\label{fig.eLR}
\end{figure}

\subsection{Eigenvalues of the strain rate tensors for $\mathsf{A}$ and $\mathsf{B}$}
Direct numerical simulations of homogeneous, isotropic turbulence show a keen tendency for the second eigenvalue of the strain rate tensor to be positive \cite{ashurst87,meneveau11}. This means that strain production (\ref{eq.R}) tends to be positive. This provides a partial explanation for the existence of the Vieillefosse tail \cite{vieillefosse84} for $\Delta \sim 0$ on the positive $R$ limb of the $Q-R$ diagram. It is convenient to examine these eigenvalues using the Lund and Rogers normalization of the second eigenvalue of the strain rate tensor, $e_{LR}$ \cite{LR94}:
\begin{equation}
e_{LR} = \frac{3\sqrt{6}R_{\mathsf{S}}}{\left(-2Q_{\mathsf{S}}\right)^{\frac{3}{2}}},
\end{equation}
where $Q_{\mathsf{S}} = -\frac{1}{2} ||\mathsf{S}||_{F}^{2}$ and $R_{\mathsf{S}} =  - \mbox{Det}(\mathsf{S})$ are the second and third invariants of the characteristic equation for $\mathsf{S}$, respectively. Hence, $-1 \le e_{LR} \le 1$ with the two end-member states pertaining to axisymmetric contractive (rod-like) and axisymmetric expansive (disc-like) conditions, respectively. The black line in Fig. \ref{fig.eLR}a highlights the well known behaviour for HIT to preferentially reside at $e_{LR} \sim 1$, meaning (because of incompressibility) that HIT favours two positive eigenvalues of similar magnitude and a negative eigenvalue of double magnitude, leading to the formation of disc-like features in HIT. These results may be contrasted with those for $e_{LR}^{(B)}$, the equivalent variable for the shear-free, normal tensor, $\mathsf{B}$ shown in Fig. \ref{fig.eLR}b, which in essence consists of peaks at $\pm 1$ with the negative peak larger than the positive peak (as a consequence, a logarithmic ordinate is adopted). More precisely, 36\% of the distribution function was found at $-1 \le e_{LR}^{(B)} \le -0.98$ while 24\% were at $0.98 \le e_{LR} \le 1$. This leads to our third result: the tendency of the strain eigenvalues to be organised such that disc-like shapes are dominant is not a consequence of strain in isolation. Without the presence of shear, turbulence is actually more likely to reside in the axisymmetric contracting regime, forming rod-like structures. This permits us to reconcile the tendency to form disc-like structures in homogenous isotropic turbulence with the preference for rod-like structure in the return to isotropy problem \cite{pope86,lumley01}, as shown recently using a novel method to establish a wide range of initial conditions \cite{li15}: A consequence of relaxing turbulence is a reduction in any mean shear and, hence, production, which reduces vorticity and induced local scale straining and shear at the small scale. Hence, the preference for rod-like structures is readily observable while, because of the masking effect of shear, this tendency is completely over-ridden in steady HIT where energy is being actively injected. The distribution of $e_{LR}^{(C)}$ in Fig. \ref{fig.eLR}c is clearly centred at zero with two equal and opposite eigenvalues for the strain rate tensor. There is, however, a slight bias in the tail towards $e_{LR}^{(C)} = +1$.


 \begin{table}
 \caption{\label{fig.eLRTable}The $3 \times 3$ classification tables for $e_{LR}$ and $e_{LR}^{(B)}$ as a function of the six regions on the $Q-R$ diagram. The emboldened number in the heading row for each sub-table is the percentage of the data found within a given region of the $Q-R$ diagram. The value in italics in the top-left panel is the single largest change from one $e_{LR}$ state to another.}
 \begin{ruledtabular}
 \begin{tabular}{|c c c| c| c c c| c|}
 $(Q > 0, R < 0)$ & \multicolumn{3}{c|}{{\bf 26}} & $(Q > 0, R > 0)$ & \multicolumn{3}{c|}{{\bf 11}}\\
\hline
 & $e_{LR} < -1/3$ & $|e_{LR}| < 1/3$ & $e_{LR} > 1/3$ & & $e_{LR} < -1/3$ & $|e_{LR}| < 1/3$ & $e_{LR} > 1/3$\\

$e_{LR}^{(B)} < -1/3$ & 14.0 & 27.5 & \emph{58.5} &  $e_{LR}^{(B)} < -1/3$ & 00.0 & 00.0 & 00.0\\
$|e_{LR}^{(B)}| < 1/3$ & 00.0 & 00.0 & 00.0 &  $e_{LR}^{(B)} < -1/3$ & 00.0 & 00.0 & 00.0\\
$e_{LR}^{(B)} > 1/3$ & 00.0 & 00.0 & 00.0 &  $e_{LR}^{(B)} < -1/3$ & 28.8 & 36.0 & 35.2\\
& & & & & & & \\
\hline
$(Q < 0, R < 0, \Delta > 0)$ & \multicolumn{3}{c|}{{\bf 10}} & $(Q < 0, R > 0, \Delta > 0)$ & \multicolumn{3}{c|}{{\bf 13}}\\
\hline
 & $e_{LR} < -1/3$ & $|e_{LR}| < 1/3$ & $e_{LR} > 1/3$ & & $e_{LR} < -1/3$ & $|e_{LR}| < 1/3$ & $e_{LR} > 1/3$\\

$e_{LR}^{(B)} < -1/3$ & 31.2 & 30.3 & 38.5 &  $e_{LR}^{(B)} < -1/3$ & 00.0 & 00.0 & 00.0\\
$|e_{LR}^{(B)}| < 1/3$ & 00.0 & 00.0 & 00.0 &  $e_{LR}^{(B)} < -1/3$ & 00.0 & 00.0 & 00.0\\
$e_{LR}^{(B)} > 1/3$ & 00.0 & 00.0 & 00.0 &  $e_{LR}^{(B)} < -1/3$ & 3.5 & 12.8 & 83.7\\
& & & & & & & \\
\hline
$(Q < 0, R < 0, \Delta < 0)$ & \multicolumn{3}{c|}{{\bf 9}} & $(Q < 0, R > 0, \Delta < 0)$ & \multicolumn{3}{c|}{{\bf 31}}\\
\hline
 & $e_{LR} < -1/3$ & $|e_{LR}| < 1/3$ & $e_{LR} > 1/3$ & & $e_{LR} < -1/3$ & $|e_{LR}| < 1/3$ & $e_{LR} > 1/3$\\

$e_{LR}^{(B)} < -1/3$ & 24.2 & 19.8 & 12.7 &  $e_{LR}^{(B)} < -1/3$ & 00.0 & 00.0 & 00.0\\
$|e_{LR}^{(B)}| < 1/3$ & 3.1 & 28.0 & 12.2 &  $e_{LR}^{(B)} < -1/3$ & 0.4 & 10.5 & 7.9\\
$e_{LR}^{(B)} > 1/3$ & 00.0 & 00.0 & 00.0 &  $e_{LR}^{(B)} < -1/3$ & 0.4 & 5.2 & 75.6\\

 \end{tabular}
 \end{ruledtabular}
 \end{table}

\subsection{Structure of the strain rate eigenvalues as a function of the $Q-R$ diagram}
Based on the preceding analysis, we form three broad groupings for $e_{LR}$ and $e_{LR}^{(B)}$: $e_{LR} < -1/3$ where there is a tendency to rod-like structures; $e_{LR} > 1/3$ where there is a tendency to disc-like structures; and, $|e_{LR}| < 1/3$ where the second eigenvector, $\boldsymbol{\lambda}_{2}$ for the second eigenvalue $\lambda_{2}$ of the strain rate tensor is close to zero. The classification of the data between these three states for $e_{LR}$ and $e_{LR}^{(B)}$ is given in Table \ref{fig.eLRTable}, subdivided by the six regions of the $Q-R$ diagram, whose relative frequency of occurrence is given by the emboldened number above the top-left of each panel. The reason for the emergence of the dominant rod-like state for $e_{LR}^{(B)}$ is made explicit here: As one would hypothesise from (\ref{eq.R}) and the sign of strain production, when $R<0$ the emergence of the rod-like state is clear. Indeed, when $\Delta>0$ all the strain rate tensors for $\mathsf{B}$ collapse onto the rod-like ($R<0$) or disc-like ($R>0$) states. It is only in the $\Delta < 0$ that a tendency emerges for $\lambda_{2} \sim 0$. Our fourth result is highlighted by the italicised value in the upper-left panel: the region of the $Q-R$ diagram most responsible for causing a change in the strain eigenvalue structure between the normal tensor, $\mathsf{B}$ and $\mathsf{A}$ is the $Q>0, R<0$ region. Here, 58.5\% of cases (15\% of the full dataset) are altered by local shear from a rod-like to a disc-like state. This result has implications for coherent structure identification algorithims, the majority of which are based on $\Delta >0$ or $Q>0$ is some way \cite{hunt88,chong90,chakraborty05}: Where enstrophy production exceeds strain production, shearing has a particularly strong impact on the nature of the structures observed. This qualitative difference between regions where $\Delta>0$ and $Q>0$ provides an additional justification for the device of expanding the $Q-R$ diagram to three dimensions based on separate analyses of strain production and enstrophy production \cite{luthi09}.

The opposite tendency to that highlighted in the $Q>0,R<0$ region, is that disc-like structures in the $Q<0, R>0$ exist as such, independent of the contribution of shear. The greater stability of the topology of the strain rate tensor in this region is very clear, with this region of the $Q-R$ diagram accommodating 44\% of the total data and over a third of the full dataset stable in the disc-like state, irrespective of the contribution from the non-normal tensor, $\mathsf{C}$. This is in accord with analysis due to Taylor and Betchov \cite{taylor37,betchov56} that enstrophy production, $\langle \boldsymbol{\omega}^{*}\mathsf{S}\boldsymbol{\omega}\rangle >0$, which means $\langle -\mbox{Det}(\mathsf{S})\rangle > 0$, with disc-like structures critical for enstrophy production. Here we have shown that this crucial role is enhanced by the reduced effects of shear in this region as seen from the $\tau_{F}$ values in the bottom right panel of Fig. \ref{fig.frobnorm}. It is also notable that the two areas of greatest interest in Table \ref{fig.eLRTable} are close to the Vieillefosse tail and in the top-left quadrant ($Q>0, R<0$). It is both of these regions where turbulence is found preferentially relative to kinematic null data \cite{tsinober01} and dynamically constrained null data \cite{k17}.

\begin{figure}
\vspace*{2mm}
\begin{center}
\includegraphics[width=8.8cm]{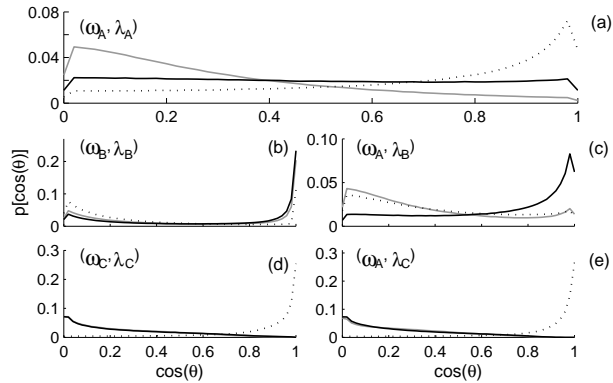}
\end{center}
\caption{Distribution functions for the cosines of the angles between the vorticity vector and strain eigenvectors for $\mathsf{A}$, $\mathsf{B}$ and $\mathsf{C}$. Given an ascending ordering of the eigenvalues of the strain rate tensors, the grey line corresponds to $\boldsymbol{\lambda}_{1}$, the dotted line to $\boldsymbol{\lambda}_{2}$ and the black line to $\boldsymbol{\lambda}_{3}$.} 
\label{fig.angvvecS}
\end{figure}

\subsection{Vorticity vector and strain rate tensor alignments in $Q-R$ space}
Defining $\theta$ as the angle between the vorticity vector, $\boldsymbol{\omega}$, and the strain rate eigenvectors, $\boldsymbol{\lambda}$ then our additive decomposition permits a richer exploration of the alignment structure of turbulence because we have three vorticity vectors and nine strain rate eigenvectors. A subset of these permutations are shown in Fig. \ref{fig.angvvecS}, with the standard decomposition in terms of $\mathsf{A}$ given in the upper panel, where the well-known preferential alignment with the second strain eigenvector, $\boldsymbol{\lambda}_{2}$, shown as a dotted line, is clear \cite{kerr85,ashurst87}. This was surprising at ht the time as it was assumed that the strongest alignment would be with the most positive strain eigenvector, $\boldsymbol{\lambda}_{3}$, shown here with a black line. Our fifth main result is clear in the comparison between Fig. \ref{fig.angvvecS}a and Fig. \ref{fig.angvvecS}c,e. The dominant alignment is actually between $\boldsymbol{\omega}_{A}$ and $\boldsymbol{\lambda}_{3}$ when the normal tensor is considered as was originally hypothesised. However, a far stronger alignment exists between $\boldsymbol{\omega}_{A}$ and $\boldsymbol{\lambda}_{2}$ when the non-normal tensor is considered, with an anti-alignment for $\boldsymbol{\lambda}_{1}$ and $\boldsymbol{\lambda}_{3}$. The results in panels (d) and (e) of Fig. \ref{fig.angvvecS} are very similar, implying a particularly close alignment between $\boldsymbol{\omega}_{A}$ and $\boldsymbol{\omega}_{C}$. With the angle between the vorticity vectors denoted by $\phi$, this was indeed found to be the case, with 60\% of data exhibiting $\mbox{cos}(\phi) > 0.94$ for these two vorticity vectors, compared to 25\% of data with $\mbox{cos}(\phi) > 0.94$ for the angles between $\boldsymbol{\omega}_{A}$ and $\boldsymbol{\omega}_{B}$, respectively.

\section{Conclusion} 
The role of small scale shear processes in homogeneous, isotropic turbulence has been previously understudied, as a consequence of the eigenvalue-based decompositions of the velocity gradient tensor that have been privileged. This paper has introduced the use of the Schur decomposition for analysing these properties, resulting in a separation of the aspects of turbulence that are attributed to eigenvalue/normality effects (e.g. the balance of enstrophy and strain, $Q$, and the balance of strain production and enstrophy production, $R$), and those that are due to non-normality/shear. We have re-examined a number of fundamental results on the structure of homogeneous, isotropic turbulence (HIT) and have found that:
\begin{itemize}
\item The size of the non-normal tensor (as measured by the Frobenius norm) is similar, on average, to the normal tensor;
\item The well-known property of the eigenvalues of the strain rate tensor to result in disc-like structures is due to the effect of the non-normal tensor - the normal tensor on its own is preferentially in a rod-like configuration;
\item The persistence of the disc-like state between the normal tensor and the velocity gradient tensor occurs preferentially close to the Vieillefosse tail ($R > 0, \Delta \sim 0$), while there is the strongest tendency for a rod-like normal tensor to change to a disc-like structure is the $R<0$, $Q>0$ region;
\item Another well known property of HIT is for the vorticity vector to be preferentially aligned with the eigenvector for the intermediate eigenvalue of the strain rate tensor. Again, we show that this is actually a property of the non-normal, shear-based tensor, with the normal tensor aligned with the eigenvector for the most positive eigenvalue (the third). 
\end{itemize} 
This final result perhaps highlights how the role of small-scale shear has not previously been integrated into turbulence analysis properly. Before the $\boldsymbol{\omega}-\boldsymbol{\lambda}_{2}$ alignment was discovered \cite{kerr85,ashurst87} it was believed that the favoured alignment should be $\boldsymbol{\omega}-\boldsymbol{\lambda}_{3}$. We have shown that the latter is true when we isolate the normal tensor. Hence, as identified twenty years ago from the perspective of the stability of hydrodynamic systems, moving beyond terms derived from eigenvalues to consider the non-normal structure permits a richer description of the dynamics \cite{schmid93,trefethen93}. This richness we have uncovered should lead to an enhanced ability to model and control turbulence.

\begin{acknowledgments}
This research was supported by a Royal Academy of Engineering/Leverhulme Trust Senior Research Fellowship LTSRF1516-12-89 awarded to the author.
\end{acknowledgments}

\bibliography{arxiv_SchurAndTurbulenceStructure}
\end{document}